# Fundamentals of impulsive energy release in the corona

Heliophysics Decadal Survey 2024–2033 white paper

Lead Author:
Albert Y. Shih, NASA Goddard Space Flight Center
albert.y.shih@nasa.gov; 0000-0001-6874-2594

Co-Authors:
L. Glesener (UMN), S. Krucker (UCB, FHNW), S. Guidoni (Amer. Univ./GSFC), S. Christe (GSFC), K. Reeves (CfA), S. Gburek (PAS), A. Caspi (SwRI), M. Alaoui (UMD), J. Allred (GSFC), M. Battaglia (FHNW), W. Baumgartner (MSFC), B. Dennis (GSFC), J. Drake (UMD), K. Goetz (UMN), L. Golub (CfA), I. Hannah (Univ. of Glasgow), L. Hayes (ESA), G. Holman (GSFC/Emeritus), A. Inglis (GSFC/CUA), J. Ireland (GSFC), G. Kerr (GSFC/CUA), J. Klimchuk (GSFC), D. McKenzie (MSFC), C. Moore (CfA), S. Musset (ESA), J. Reep (NRL), D. Ryan (FHNW), P. Saint-Hilaire (UCB), S. Savage (MSFC), D. B. Seaton (SwRI), M. Stęślicki (PAS), T. Woods (Univ. of Colorado/LASP)

Synopsis:
It is essential that there be coordinated and co-optimized observations in X-rays, gamma-rays, and EUV during the peak of solar cycle 26 (~2036) to significantly advance our understanding of impulsive energy release in the corona.  The open questions include: What are the physical origins of space-weather events? How are particles accelerated at the Sun? How is impulsively released energy transported throughout the solar atmosphere? How is the solar corona heated? Many of the processes involved in triggering, driving, and sustaining solar eruptive events—including magnetic reconnection, particle acceleration, plasma heating, and energy transport in magnetized plasmas—also play important roles in phenomena throughout the Universe.  This set of observations can be achieved through a single flagship mission or, with foreplanning, through a combination of major missions (e.g., the previously proposed *FIERCE* mission concept).

# Introduction

Solar eruptive events are the most energetic and geo-effective space-weather drivers. They originate in the corona near the Sun's surface as a combination of solar flares (impulsive bursts of radiation across the entire electromagnetic spectrum) and coronal mass ejections (CMEs; expulsions of magnetized plasma into interplanetary space). The radiation and energetic particles they produce can damage satellites, disrupt telecommunications and GPS navigation, and endanger astronauts in space.

Many of the processes involved in triggering, driving, and sustaining solar eruptive events—including magnetic reconnection, particle acceleration, plasma heating, and energy transport in magnetized plasmas—also play important roles in phenomena throughout the Universe, such as in magnetospheric substorms, gamma-ray bursts, and accretion disks. The Sun is a unique laboratory to better understand these fundamental physical processes.

We outline four interrelated areas of science investigation that would significantly advance our understanding of impulsive energy release in the corona:
- What are the physical origins of space-weather events?
- How are particles accelerated at the Sun?
- How is impulsively released energy transported throughout the solar atmosphere?
- How is the solar corona heated?

# What are the physical origins of space-weather events?

What processes trigger solar eruptions and the release of particles into the heliosphere is a long-standing debate. Understanding their origins is not only a primary goal in Heliophysics, but is indispensable for making the reliable space-weather predictions needed for future space exploration, including NASA's Artemis lunar program, as well as the National Space Weather Strategy and Action Plan. Current predictive capabilities are insufficient, in large part because of the lack of observations necessary to understand the fundamental processes underlying eruptions and their propagation into the heliosphere to form space weather.

Most models for the initiation of solar eruptive events involve magnetic reconnection, which results in both plasma heating and particle acceleration in the vicinity of the reconnected field lines. However, different models of CME initiation predict observationally differentiable timings and locations for the reconnection process, and consequently for the observational signatures (Figure 1). The internal-tether-cutting model (e.g., [1]) predicts that reconnection occurs below the flux rope before the fast takeoff of the eruption, the breakout model [2] predicts that reconnection occurs above the flux rope before fast takeoff, and the ideal MHD instability model [3] predicts that the flux rope begins to rise before reconnection occurs in either place.

**The inability of past and current instruments to image the faint signatures of accelerated particles and heated plasma in the neighborhood of (in particular, above) the main energy-release site has made it impossible to distinguish among models for the initiation of solar eruptive events.**

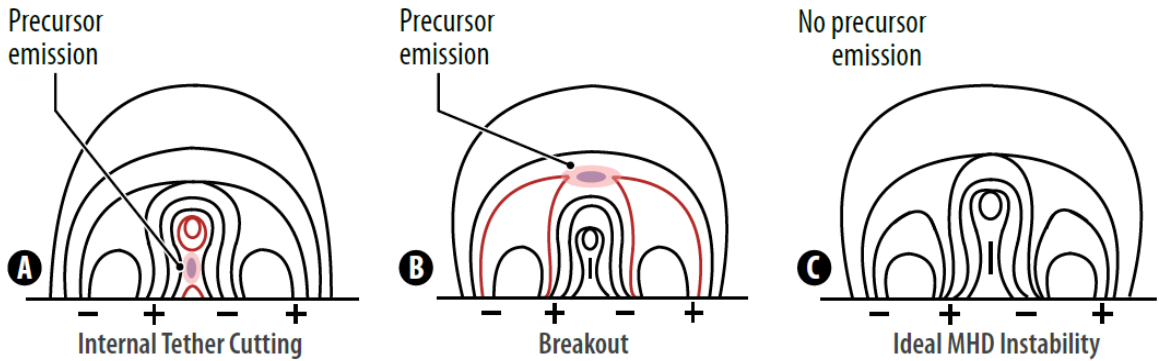

*Figure 1: Different models of the initiation of solar eruptive events can be distinguished during the early rise phase of the eruption by the differences in the timing and locations of X-ray sources from accelerated electrons and X-ray and extreme-ultraviolet (EUV) emission from heated plasma. The cartoons of possible magnetic configurations are adapted from [4].*

## How are particles accelerated at the Sun?

Accelerated particles constitute a significant fraction (up to tens of percent) of the magnetic energy released in the most eruptive space-weather events (e.g., [5]). The two most credible models explaining such efficient acceleration both involve magnetic reconnection in a current sheet below a rising flux rope that eventually becomes the CME. The first model uses a first-order Fermi acceleration process via evolving and merging "magnetic islands" [6] created by the reconnection. The second model uses a stochastic, or second-order Fermi, acceleration process in the turbulent plasma of the reconnection outflow jets (e.g., [7]). **Distinguishing between these models requires observations of the locations, sequencing, emission timescales, and number of accelerated electrons in multiple thermal and nonthermal coronal X-ray sources in the presence of intense footpoint emissions.**

There are additional science questions associated with the acceleration of ions, which have traditionally been more challenging than electrons to observe. See the white paper "Ion acceleration in solar eruptive events" by Shih et al. for detailed discussion about ion science objectives and required measurements.

## How is impulsively released energy transported throughout the solar atmosphere?

As electrons exit the coronal acceleration region and propagate along loops, they lose energy through Coulomb collisions with ambient particles, wave-particle interactions, and the generation of return currents. These energy-transport processes link the impulsively released magnetic energy to the radiation associated with flares and space-weather phenomena. Understanding this link and the relative importance of each process requires measuring hitherto poorly constrained observational inputs (e.g., the low energy cutoff) to the latest state-of-the-art numerical solar flare models (e.g., [8][9]). The resulting predictions must then be compared to previously unachievable observations, including the number and spectrum of the accelerated electrons both at the acceleration region and various locations along their paths.

# How is the solar corona heated?

A long-standing enigma in solar and stellar physics is how the atmosphere can be orders of magnitude hotter than the surface of the star itself.  This temperature difference requires some form of non-radiative heating (e.g., wave dissipation), Joule heating, or small-scale magnetic dissipation (e.g., in "nanoflares").  Both observation and theory indicate that the non-flaring corona is heated in an impulsive fashion (e.g., [10]).  Whether the dominant mechanism involved is magnetic reconnection or MHD wave dissipation has not been established.

First, we must determine if the characteristics of energy release in the smallest detectable events are fundamentally different from those in larger flares.  The number of flares as a function of their thermal energy follows a power-law over several orders of magnitude (e.g., [11]), suggesting that the underlying energy-release process scales similarly.  However, it is unknown whether nanoflares are part of this distribution.

Second, we must infer the properties of heating events too small to be detected individually, including indirectly determining the frequency with which they occur.  The presence of temperatures exceeding 5 MK in non-flaring active regions is often called the "smoking gun" of impulsive, low-frequency nanoflare heating.  Steady or high-frequency wave heating cannot maintain such high temperatures without violating other observational constraints (e.g., [12]).  The distribution of frequencies also provides valuable information on the, as yet, unestablished sequence of events leading up to reconnection.

# Required observations

The observations required for these science investigations are high-resolution imaging spectroscopy, with high sensitivity, of the following types:
- X-rays and gamma rays for energetic electrons and ions
- X-rays and EUV optimized for hot flare plasma
- EUV optimized for the dynamic evolution of flare magnetic structures

See the white paper "Solar flare energy partitioning and transport: the impulsive phase" by Kerr et al. for detailed discussion on measurement requirements.  Missions that are currently operating or are planned do not make the measurements in sufficient detail to make significant progress in any of the above science investigations.  This measurement deficiency means that solar cycle 25 will unfortunately be a missed opportunity.  **It is essential that there be coordinated and co-optimized high-energy observations during the peak of solar cycle 26 (~2036).**

While this set of observations could be realized by a single flagship mission (see the white paper by Caspi et al. for the *COMPLETE* mission concept), it could alternatively be realized through a combination of multiple major missions, as long as the coordination and co-optimization is expressly planned from the outset.  The *Fundamentals of Impulsive Energy Release in the Corona Explorer* (*FIERCE*) mission concept [13] would provide co-optimized X-ray and EUV observations within the typical schedule and cost cap (~4 years to launch and ~$250 million) for a NASA Medium-Class Explorer (MIDEX) mission.  *FIERCE* combines an X-ray spectroscopic imager that uses focusing optics to achieve high sensitivity and high dynamic range with a fast-cadence EUV imager that avoids saturation even when imaging intense flares, with no new technology development needed for the instruments.  *FIERCE* could then be complemented by next-generation gamma-ray instrumentation, perhaps even flown on balloons.

# References

1. Moore, R. L., et al. (2001), ApJ, 552, 833. [DOI:10.1086/320559]
2. Antiochos, S. K., et al. (1999), ApJ, 510, 485. [DOI:10.1086/306563]
3. Török, T., & Kliem, B. (2005), ApJ, 630, L97. [DOI:10.1086/462412]
4. Chifor, C., et al. (2007), A&A, 472, 967. [DOI:10.1051/0004-6361:20077771]
5. Emslie, A. G., et al. (2012), ApJ, 759, 71. [DOI:10.1088/0004-637X/759/1/71]
6. Drake, J. F., et al. (2006), Nature, 443, 553-556. [DOI:10.1038/nature05116]
7. Miller, J. A., et al. (1997), JGR, 102, 14631. [DOI:10.1029/97JA00976]
8. Allred, J. C., et al. (2015), ApJ, 809, 104. [DOI:10.1088/0004-637X/809/1/104]
9. Reep, J. W., et al. (2019), ApJ, 871, 18. [DOI:10.3847/1538-4357/aaf580]
10. Klimchuk, J. A. (2015), Philosophical Transactions of the Royal Society of London Series A, 373, 20140256. [DOI:10.1098/rsta.2014.0256]
11. Hannah, I. G., et al. (2011), Space Science Reviews, 159, 263. [DOI:10.1007/s11214-010-9705-4]
12. Schmelz, J. T., et al. (2009), ApJ, 693, L131. [DOI:10.1088/0004-637X/693/2/L131]
13. Shih, A. Y., et al. (2020), AMS 100th Annual Meeting, Boston, MA. [DOI:10.5281/zenodo.3674079]